\documentstyle[epsfig]{aipproc}
\begin{document}
\title{High Energy Neutrino Astronomy\\
--the cosmic-ray connection\thanks{Research supported in part
by NASA grant NAG5-7009}}
\author{Thomas K. Gaisser}
\address{Bartol Research Institute, University of Delaware\\
Newark, DE 19716}


\maketitle
\begin{abstract}
Several of the models for origin of the highest energy cosmic
rays also predict significant neutrino fluxes.  A common factor
of the models is that they must provide sufficient power to
supply the observed energy in the extragalactic component
of the cosmic radiation.  The assumption
that a comparable amount of energy goes into high-energy neutrinos allows
a model-independent estimate of the neutrino signal that may be expected.
\end{abstract}

\section{Introduction}
An important argument in favor of supernova explosions as the power
source for galactic cosmic rays is the fact that kinetic energy
of the ejecta supplies the right amount of power.  The energy
content of the cosmic radiation is
\begin{equation}
\rho_E = {4\pi\over c}\,\int\,E\,\phi(E)\,dE\;\approx\;10^{-12}\;
{\rm erg/cm}^3,
\label{Edensity}
\end{equation}
where $\phi(E)=dN/dE$ is the measured local flux of cosmic rays corrected
for the effect of solar modulation.
In the source region the average energy density in
cosmic rays is related to the average production rate per unit volume,
$q(E)$, by
\begin{equation}
\rho_E\;=\;q(E)\times\tau_{esc}(E),
\label{CRsource}
\end{equation}
where $\tau_{esc}$ is the characteristic residence time
of cosmic rays in the source region (for example the disk of
the galaxy).  The characteristic time, $\tau_{esc}(E)$,
decreases with energy \cite{Garcia-Munoz} so that the
observed spectrum is somewhat steeper than the source spectrum.

Given an estimate of $\tau_{esc}$ and the rate of
supernova explosions, it is possible to estimate the fraction
of energy of supernova explosions needed to maintain the
galactic cosmic rays in steady state, assuming that supernovae
provide the power, $q(E)$.  Assuming a rate of
three supernovae per century with a kinetic energy of $10^{51}$ ergs
per supernova, the conclusion is that an efficiency of $\sim 10$\%
for conversion of kinetic energy of supernova ejecta into
relativistic cosmic rays would suffice.
In the years since this coincidence was pointed out \cite{Ginzburg},
a theory of cosmic-ray modified shocks with a high
efficiency for particle acceleration has been developed \cite{crmodified}.
A compelling feature of this theory is that it produces
a spectral index that fits in well with what is
observed after energy dependence of propagation is accounted for.

Here we want to apply a similar analysis of energetics to cosmic rays
of extragalactic origin  The reason to focus on the high energy
end of the cosmic ray spectrum in connection with high energy
neutrino astronomy is that several models have been suggested
as sources of ultra-high energy cosmic rays which would also be likely
sources of high energy neutrinos.  These include some models of active
galactic nuclei (AGN) and some models of gamma-ray burst (GRB) sources.

In the case of galactic cosmic rays, if the production rate of neutrinos
is proportional to that of cosmic rays, $q_\nu(E)\,=\,f\times q(E)$,
we would expect a
flux of neutrinos related to the cosmic-ray flux by
\begin{equation}
\phi_\nu \propto {V_D\over c}\times f\times \phi_{cr},
\label{CRneutrino}
\end{equation}
where $f$ is the efficiency with which the cosmic rays interact
to produce neutrinos (either in the source region or in the interstellar
medium) and $V_D$ is the velocity of diffusion of cosmic rays out of the galaxy.
$V_D\sim R_{galaxy}/\tau_{esc}$ depends on the model of cosmic-ray propagation
in the galaxy, but in any case $V_D\ll c$.  Thus the galactic neutrino
flux should be suppressed by a large factor relative to the parent
cosmic rays because of their straight-line propagation out of the galaxy.
For the same reason, neutrinos, like gamma-rays, would be expected to show
the structure of the galactic disk.

For a cosmological distribution of sources of ultra-high energy
cosmic rays we assume initially, for the rough estimates discussed below, that
\begin{equation}
V_D\;\sim\;{R_H/\tau_H}\;\sim\;{3000\,{\rm Mpc}\over 10^{10} yrs}\;\approx\;c.
\label{Hubble}
\end{equation}
This approximation neglects evolution and
assumes that the intergalactic magnetic fields are weak enough so that 
charged particles can reach us from distant sources in less than
the age of the universe.  The situation for super-GZK particles, where
the maximum distance of propagation is limited to $R\ll R_H$ by
photo-pion production, needs a separate discussion.  The possibility
of large inter-galactic magnetic fields and local concentrations
of sources \cite{local1,local2}
would also have to be considered for a full treatment of
the cosmic-ray spectrum.  We use the approximation of Eq.~\ref{Hubble}
primarily to derive an estimate of the diffuse flux of high energy
neutrinos that may be associated with the sources of extragalactic
cosmic rays.

The outline of the paper is as follows.  First we discuss the
transition from galactic to extragalactic cosmic radiation in order
to define an extragalactic component.  Then we compare the power needed
for the extragalactic cosmic radiation with that available from
various potential sources.  We conclude with a review of the
predictions for detection of high energy neutrinos.

\section{Extragalactic component of cosmic rays}
It is generally believed that sources of the highest energy
cosmic rays are extragalactic, or at least not confined
to the plane of the galaxy.  Indeed, there is some evidence for a
transition from one particle population
to another somewhere above $10^{18}$~eV.  There is a trend from heavy
toward lighter composition in the measurements of $X_{\rm max}$
versus energy (see Fig. 1) and a suggestion of a hardening of the spectrum
between $10^{18}$ and $10^{19}$~eV (see Fig. 2).  The picture
is not as clear as it first appeared in the original stereo Fly's
Eye result~\cite{FEstereo}.  For example, the coincident measurements
of the prototype HiRes Fly's Eye with the MIA ground array \cite{MIAHiRes},
as shown by the gray, filled circles in Fig.~\ref{Fig2}, have a steeper
slope and trend toward the proton curve more quickly than the
original Fly's Eye data (black triangles).  This would 
indicate a stronger change of composition at a somewhat lower energy.

\begin{figure}[htb]
\flushleft{\epsfig{figure=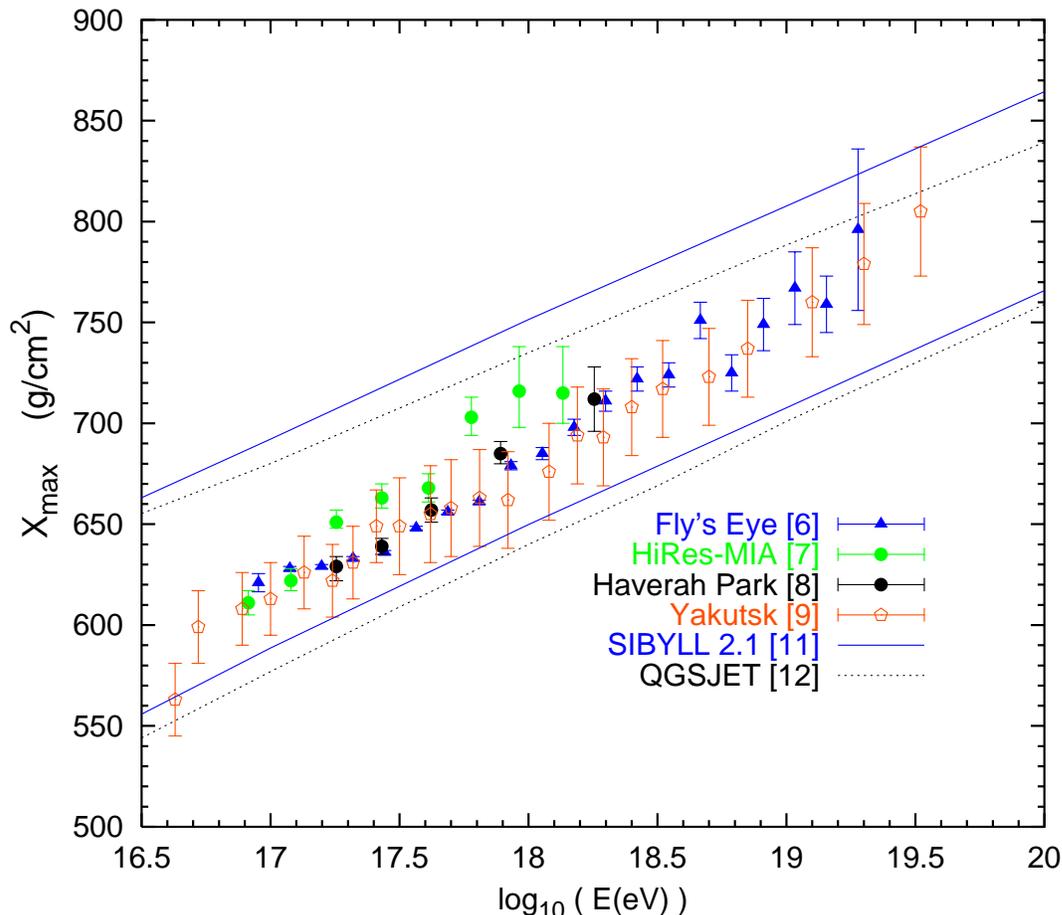,width=14.5cm}}
\caption{Average depth of shower maximum (\protect$X_{max}$) vs. 
energy~\protect\cite{FEstereo,MIAHiRes,Hinton,Yakutsk}
compared to calculated~\protect\cite{Pryke} 
protons (upper curves) and iron primaries (lower curves)
in two models~\protect\cite{SIBYLL21,QGSjet}. 
}
\label{Fig1}
\end{figure}
Another relevant piece of information is the anisotropy measurements from
AGASA~\cite{AGASAanisotropy} and Fly's Eye~\cite{FEanisotropy},
which can be interpreted as an
enhancement of particles from near the direction of the
galactic center in the energy bin around $10^{18}$~eV~\cite{Clay2}.
The anisotropy disappears at higher energy.

Although the experimental picture is still not entirely clear,
in order to estimate the power 0 we assume in what follows
that the transition to cosmic rays of extragalactic origin occurs
between $10^{18}$ and $10^{19}$~eV. 

\section{Power for extragalactic cosmic rays}
Measurements of the cosmic-ray spectrum above $10^{17}$~eV are
summarized in Fig. 2.  There are at least three
problems that must be dealt with to estimate the power required
to supply the extragalactic cosmic radiation.  First is where the
transition to the extragalactic component occurs in the data.
As discussed above, we assume that the particles with $E>3\times10^{18}$~eV
are mostly of extragalactic origin and normalize the
extragalactic component at $E\ge10^{19}$~eV.  To set the scale for the uncertainty
in this assumption, we investigate below
the consequences of increasing or decreasing
the crossover energy by half a decade.
The second problem is to decide how to extrapolate to lower energy
where the observed spectrum is likely dominated by cosmic rays
from inside the galaxy.  This is important
because most of the energy content is likely to be in the lower energy
particles.  Finally, we must deal separately with the super-GZK particles
whose sources must be relatively nearby.  We deal with these points
in turn.

\begin{figure}[htb]
\flushleft{\epsfig{figure=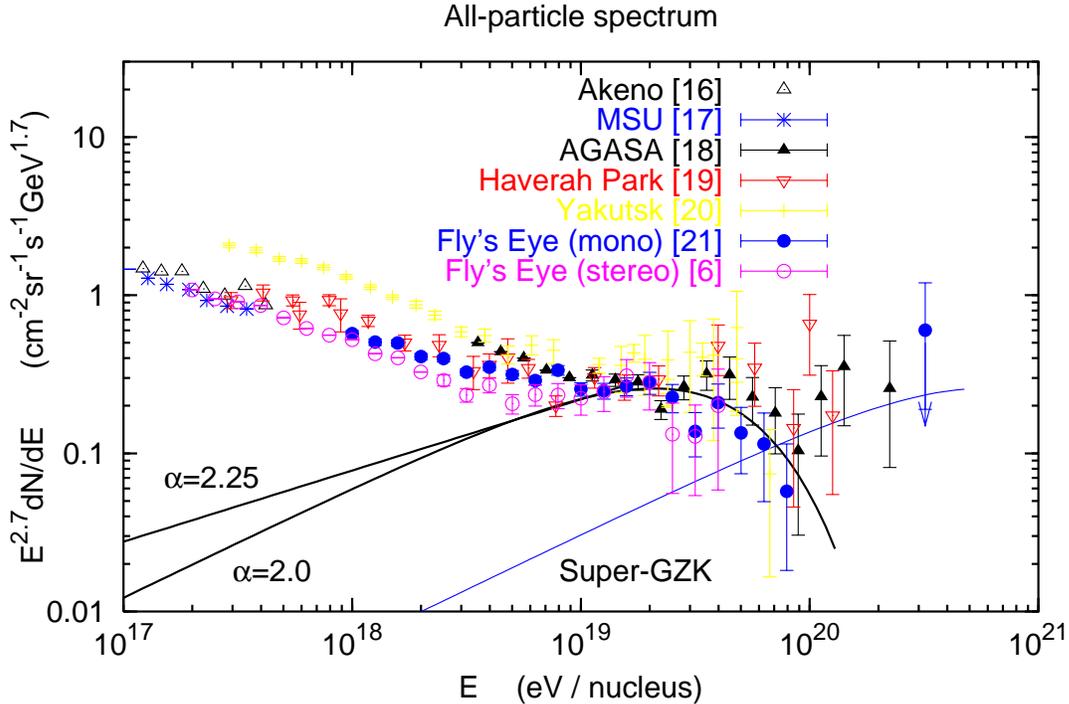,width=14.5cm}}
\caption{The high energy cosmic rays 
spectrum.  See text for explanation of curves.  Data are
from Refs.~\protect\cite{Akeno,MSU,AGASA,HP,Yakutsk2,FEmono,FEstereo}}
\label{Fig2}
\end{figure}
\subsection{Normalization at high energy}
Normalizing an $E^{-2}$ differential spectrum, as is generally
assumed for the sources of the highest energy cosmic
rays~\cite{Waxman,Piran},
to a measured spectrum with a differential spectral index close to
$\alpha\,=\,3$ is a highly uncertain operation.  The lower heavy
line in Fig. 2 shows a spectrum with $\alpha = 2$ and an exponential
cutoff at $5\times10^{19}$~eV to represent the GZK effect.  The
excess of data above the curve for $E<10^{19}$~eV may be attributed
to the high energy tail of the galactic cosmic-ray spectrum.
Another possibility could be that we live inside a 
cosmologically local concentration of extragalactic sources
of cosmic rays which propagate with an energy-dependent $\tau_{esc}$
analogously to galactic cosmic rays.

Integrating the energy content under this curve
as in Eq.~\ref{Edensity} gives for the energy density in
cosmic rays of extragalactic origin,
$\rho_{EG} \sim 2\times 10^{-19}$~erg/cm$^3$.
Note that this result assumes the extragalactic spectrum
extends down to $\sim 1$~GeV.
 Replacing $\tau_{esc}\rightarrow \tau_H\approx 10^{10}$~yrs
in Eq.~\ref{CRsource} then leads to the estimate
$q_{EG}\sim 10^{37}$~erg/Mpc$^3$/s.  Shifting the
normalization point lower (or higher) by half a
decade in energy would increase (decrease) this estimate by
roughly a factor of two.  This is comparable to the systematic
differences among the different measurements of the spectrum.

{\large
\begin{table}
\caption{Power per source for extragalactic cosmic rays}
\center{
\begin{tabular}{||ll||}  \hline
Source density (rate) & Power per source \\ \hline
 & \\
$3\times 10^{-3}$ galaxies/Mpc$^3$ & $3\times 10^{39}$ erg/s/galaxy \\
$3\times 10^{-6}$ clusters/Mpc$^3$ & $3\times 10^{42}$ erg/s/cluster \\
$10^{-7}$ AGN/Mpc$^3$ & $10^{44}$~erg/s/Active galaxy \\
$1000$ GRB/yr & $3\times 10^{52}$~erg/GRB \\ \hline
\end{tabular}
}
\end{table}
}

Table 1 shows what this power requirement would imply
for various classes of potential sources.  
In each case, these are comparable to observed
luminosities.  Therefore
all are plausible potential sources provided a mechanism exists
to achieve $E_{max}\sim10^{20}$~eV.
(For clusters of galaxies see Refs.~\cite{Norman,Kang}; for
active galaxies~\cite{AGN} and for GRB~\cite{WB,Vietri}.)

\subsection{Extrapolation to low energy}
If the spectral index of the extragalactic source spectrum is
steeper than $\alpha = 2$, then the power requirement will be
greater.  Both AGN and (especially) GRB involve relativistic
shocks.  Acceleration at relativistic shocks typically produces
a spectral index $\alpha\approx 2.25$~\cite{Achterberg,Ostrowski,Kirk}.
The difference is particularly important in the case of GRB
where the bulk Lorentz factor $\Gamma\sim 300$.  Such a spectrum,
with the same normalization at $10^{19}$~eV is shown by the
upper heavy line in Fig. 2.  In general
\begin{equation}
{q(\alpha)\over q(\alpha=2)}\;=
  \;\left({10^{19}\,{\rm eV}\over E_{min}}\right)^{(\alpha-2)}
\times {1\over (\alpha - 2)\,\ln(E_{max}/E_{min})},
\label{enhancement}
\end{equation}
where the normalization is fixed at $10^{19}$~eV.  For $\alpha \approx 2.3$
and $E_{min}\approx 1$~GeV, the power requirement is a factor of $\sim100$
greater than for $\alpha=2.0$.

Vietri \cite{Vietri2} argues, however, that in the case of a
relativistic shock accelerating particles from swept up material,
$E_{min}\sim\Gamma^2\times m_p$,
which corresponds to $E_{min}\sim100$~TeV for GRB with $\Gamma\sim300$.
In this case the enhancement factor in Eq.~\ref{enhancement} is only
about a factor of 3, and the power requirement increases from
$3\times 10^{52}$ to $10^{53}$~erg/GRB.

\subsection{Super-GZK particles}
One possibility is that the particles above the GZK cutoff
may be due to a local concentration of the same type of
sources that produce a universal component \cite{BW,Grigorieva}.
The line labelled {\it Super-GZK} in Fig. 2
represents a possible contribution
from such nearby sources, assuming $\alpha=2$.
The energy integral for this component 
is approximately a factor of four lower than for the corresponding
universal contribution, but the power density to supply it is higher
by a factor of five or ten because the distance from which
sources can contribute is limited by energy loss due to
photo-pion production.  A factor of 5-10 local overdensity
of sources of ultra-high energy cosmic rays is difficult
to reconcile with other considerations.  For example,
Ref.~\cite{Stecker} points out that most of the GRB rate comes from
sources with $z\sim 1$.  (See Ref.~\cite{Dermer} for a
different GRB scenario.)    A local
overdensity of more than a factor $\sim2$ would also be difficult to
reconcile with data on the large scale distribution of matter in
our vicinity of the universe\cite{Blasi}.  Given the limited
statistics, however, this explanation cannot be ruled out if
the source spectrum is hard enough ($\alpha\le 2$~\cite{Blasi}).

Another possibility is that the super-GZK particles have a different
origin altogether, being products of parton cascades generated by decay or
annihilation of GUT-scale objects, such as topological defects~\cite{TD}
or massive relic particles~\cite{Berezinsky}.  In both cases the cascade
consists of hadronization of partons at extremely high mass scale.
Thus the ratio of photons and neutrinos to protons at production is
large because pions dominate the hadronic cascade.  However, in the
case of a cosmological distribution of sources, the photons will
initiate electromagnetic 
cascades during propagation in the microwave background, reducing
their contribution to the observed super-GZK events~\cite{Bhat}.  In the model
of decaying massive relics, however, the predominant contribution
comes from relatively nearby -- particles concentrated
in the halo of the galaxy.  In this
case, electromagnetic cascading will be negligible, 
and the highest energy events
would, for the most part, have to be photon-initiated showers.
It has been known for some time that the big Fly's Eye event~\cite{FEbig},
for which the profile is measured, looks more like a shower
initiated by a proton or nucleus than a photon~\cite{HSV}.
Recently, an analysis of horizontal air showers~\cite{Ave}
showed that most of the observed events above the GZK cutoff cannot
be initiated by photons.

\section{Expected neutrino fluxes}
A standard technique to search for high energy neutrinos of astrophysical
origin is to look for upward-moving muons induced by $\nu_\mu$ that
have penetrated the Earth.  The signal is the convolution
\begin{equation}
Signal\;\sim\;Area\otimes R_\mu\,N_A\otimes \sigma_\nu\otimes \phi_\nu,
\label{convolution}
\end{equation}
where $R_\mu$ is the muon range in g/cm$^2$ and $N_A$ is Avogadro's number.
The range and cross section both increase linearly with energy
into the TeV region, after which the rate of increase slows.
Neutrinos with $E_\nu < 100$~TeV are not strongly attenuated
by the Earth, and much of the solid angle away
from the nadir remains accessible up to 1~PeV~\cite{Quigg}.
Thus the optimum range for $\nu_\mu$-induced upward muons
is from a TeV to a PeV.  Also in this energy range the
muon energy loss is greater than minimum ionizing, which
is a potential way to discriminate against the background
of atmospheric neutrinos, which have a steeply falling spectrum.
In what follows, I use the cross sections from Ref.~\cite{Quigg},
taking account of absorption by the Earth, to estimate $\nu_\mu$-induced
signals in a kilometer-scale detector.  A minimum pathlength
of $0.5$~km inside the detector is required, which corresponds to
a threshold of approximately $E_\mu\ge100$~GeV in water.

\subsection{Generic estimate}
Using the normalization of \S~III-A and assuming $\alpha\sim2.0$ for the
neutrinos as well as the cosmic rays, one estimates a signal
of $f\times 30$ events/km$^2$/yr~\cite{Taormina}, where $f$ is the efficiency
for production of neutrinos relative to cosmic rays.
For $f=1$ this is essentially the 
Waxman-Bahcall upper bound~\cite{Waxman},
which applies for sources that are transparent to neutrons.  If
evolution is included, this estimate may be increased by\
factor of about five
if sources are assumed to evolve similarly to the
rate of star formation~\cite{Waxman,MPR}.  This is because the ultra-high energy
protons from high redshift would be attenuated by photoproduction
and pair production while the neutrinos would not.  Thus
the neutrino flux would be greater for a given cosmic-ray flux
at the normalization point.  In addition, if the spectrum of extragalactic
cosmic rays to which the normalization is made has $\alpha>2.0$,
then the corresponding estimated neutrino flux in the TeV to
PeV range and the corresponding signal could also
be larger.

\begin{figure}[htb]
\flushleft{\epsfig{figure=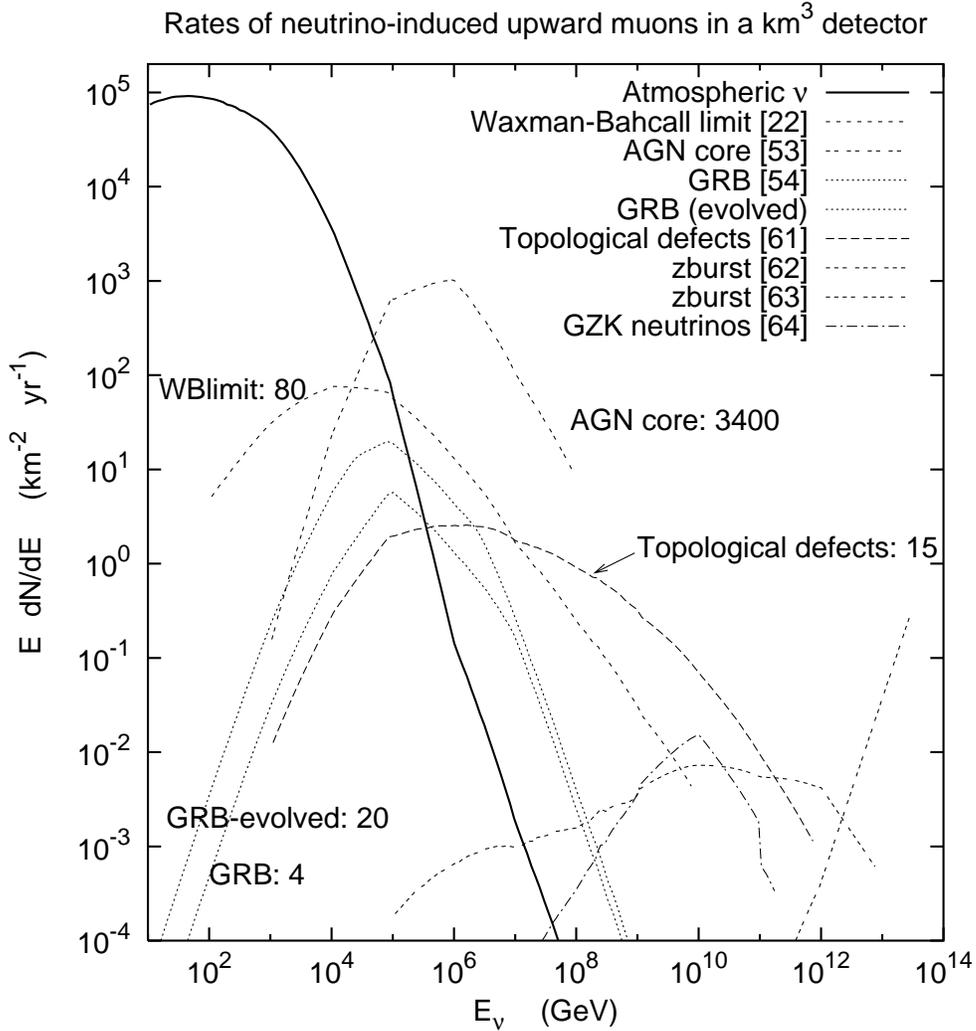,width=14cm}}
\caption{Summary of predicted neutrino signals in a kilometer-scale
detector.  The diffuse signal per year per steradian per
logarithmic energy interval is plotted
as a function of \protect$E_\nu$.  Numbers indicate events per year
above atmospheric background (see text).}
\label{Fig3}
\end{figure}
\subsection{Predictions of models}

Fig. 3 shows the rates of neutrino-induced upward muons predicted
in various specific models.  These are diffuse fluxes summed over
$2\pi$~sr with absorption in the Earth accounted for.  The
background induced by atmospheric $\nu_\mu$ is shown as the heavy
solid line.  The flattening of the atmospheric background at 1 PeV
is due to an assumed prompt component of neutrinos from charm
decay~\cite{charm}, the level of which is rather uncertain~\cite{charm2}.
The numbers indicate events per year above atmospheric background,
which I have estimated by taking the integral of the signal flux
above the energy where it crosses the atmospheric background.
This crossover is generally in the range of 100 -- 300 TeV.
Present upper limits from
Frejus~\cite{Frejus}, Baikal~\cite{Baikal} and AMANDA~\cite{AMANDA}
rule out one of the original AGN core models~\cite{Protheroe},
and the model of Ref.~\cite{Stecker2} (shown as AGN core in Fig.~\ref{Fig3})
is marginally allowed~\cite{Frejus}
or marginally ruled out~\cite{AMANDA}.
The GRB curves in Fig.~\ref{Fig3} are calculated from Eq.~\ref{convolution}
 starting 
from the unevolved GRB neutrino flux 
plotted in Ref.~\cite{Waxman}.  I assumed source evolution from
Ref.~\cite{Waxman} to obtain the GRB-evolved neutrino spectrum.

The GRB model~\cite{WB2}, has an interesting feature, which
illustrates the important point that the neutrino spectrum need
not have the same shape as the parent spectrum of accelerated
protons~\cite{RM}.  In this case the target for pion production is assumed
to be the X-ray/$\gamma$-ray photons
in the expanding fireball, which have approximately
an $E_\gamma^{-2}$ differential spectrum above a characteristic energy
$\epsilon_b$ and an $E_\gamma^{-1}$ spectrum below the break.  
For protons with energy sufficiently high
to photoproduce on photons with $E_\gamma<\epsilon_b$, the resulting
neutrino spectrum follows the parent proton spectrum.  For lower
energy protons, the density of target photons above threshold for
photo-pion production decreases as the proton energy decreases.  The
result of this convolution is a break (steepening) in the
neutrino spectrum, which, for the parameters of Ref.~\cite{WB2}
occurs at $E_\nu\sim 100$~TeV.
Since both $R_\mu$ and $\sigma_\nu$
are increasing with energy, this is where the signal peaks.  

The key to detecting a diffuse signal above the atmospheric 
background is to look for an excess of events of high energy.
Typically, muons with energies of several TeV and higher
will radiate one or two bursts per kilometer of water in
which they deposit some 10\% of their energy.  Such bursting tracks
may be a useful signature.

\subsection{Point sources}
With point sources, the atmospheric background can be reduced
to an extent that depends on the angular resolution of the
detector.  As an example, Fig. 4 shows the signal that would
be generated by a neutrino flux normalized to the level of the TeV
$\gamma$-ray emission during the extended
high-state of Mrk501 in 1997~\cite{Mrk1,Mrk2,Mrk3},
assuming $\phi_{\nu_\mu+\bar{\nu}_\mu}\,=\,\phi_\gamma$,
(specifically using the fit of Ref.~\cite{Mrk1}).
The atmospheric 
background is calculated assuming that the detector has a pointing
resolution of $1^\circ$ space angle, and the angle between
the detected muon and the neutrino that produced it has been
accounted for.  The integral of the signal shown in the figure
would give $\sim30$~events per year above atmospheric background.
Unfortunately, Mrk501 is not normally
such an intense source~\cite{HEGRA} as it was for half of 1997.

\begin{figure}[htb]
\flushleft{\epsfig{figure=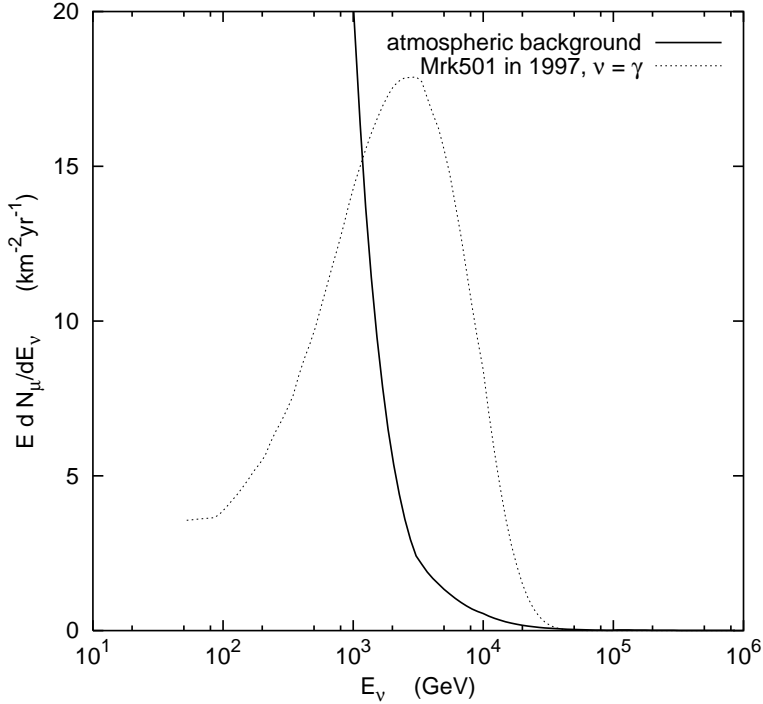,width=10cm}}
\caption{Predicted signal of $\nu$-induced muons from a
source with \protect$\phi_\nu = \phi_\gamma$(Mrk 501, high).  (See text.)}
\label{Fig4}
\end{figure}

It is interesting that there is at least one AGN-blazar model
in which a neutrino flux comparable to the $\gamma$-ray flux
would be expected~\cite{Pohl}.  In this model the high energy
photons and neutrinos come from decay of pions produced when
blobs of ultra-relativistic gas collide with the interstellar
medium of the host galaxy near the central engine of the AGN.
Farther out the material slows down and merges into the jets
that extend to large distance.

\subsection{Ultra-high energy neutrinos}
Some of the ``top-down" models for the ultra-high energy cosmic
rays~\cite{Sigl,Gelmini,Yoshida} 
predict an intensity of neutrinos with energies
in the PeV range and higher that is somewhat above the neutrinos
photo-produced by ultra-high energy cosmic rays on the microwave
background (GZK neutrinos).  The rates of $\nu_\mu$-induced
upward neutrinos shown for those models in Fig.~\ref{Fig3} are strongly
attenuated by absorption in the Earth  
(although regenerated $\nu_\tau$ could emerge with
degraded energy~\cite{NuTau}).  However, the energy is high
enough so that charged-current interactions of $\nu_e$ from
above could be identified by the large electromagnetic cascades
they would produce inside a kilometer-scale detector~\cite{Nuebursts}.
For example, the rate of downward events predicted by the model from
Ref.~\cite{Yoshida} would be $\sim10$ per year per km$^3$
even though the integrated rate of upward $\nu_\mu$-induced muons shown
in Fig.~\ref{Fig3} is $\ll1$.

\subsection{Concluding comment}
Estimates of the type discussed here -- either based on
energetics and a generic relation to extra-galactic cosmic rays,
or on specific models -- set the scale for a high energy
neutrino telescope.  The estimates are of the order of tens of events per year
in a kilometer-scale detector.
While the hope is that new kinds of 
``hidden'' neutrino sources~\cite{hidden} will be discovered,
the detector should be designed to find signals from likely
known sources at this level.  
\vspace{0.3cm}

\noindent
ACKNOWLEDGEMENTS.  I am grateful for helpful conversations with 
Jaime Alvarez-Mu\~{n}iz, Todor Stanev and Mario Vietri.

\end{document}